\documentclass[letterpaper, conference]{ieeeconf}

\IEEEoverridecommandlockouts                              
\usepackage{url} 
\usepackage{graphicx}
\usepackage{float}
\usepackage{algorithm}
\usepackage{algorithmic}
\usepackage{tablefootnote}
\title{\LARGE \bf Preventing Poisoning Attacks on AI based Threat Intelligence Systems}

\author{Nitika Khurana, Sudip Mittal and Anupam Joshi \\
University of Maryland, Baltimore County, Baltimore, MD 21250, USA\\
Email: $\lbrace$nkhur1, smittal1, joshi$\rbrace$@umbc.edu
}

\begin{document}

\maketitle
\thispagestyle{empty}
\pagestyle{empty}

\begin{abstract}

As AI systems become more ubiquitous, securing them becomes an emerging challenge. Over the years, with the surge in online social media use and the data available for analysis, AI systems have been built to extract, represent and use this information.  The credibility of this information extracted from open sources, however, can often be questionable. Malicious or incorrect information  can cause a loss of money, reputation, and resources; and in certain situations, pose a threat to human life. In this paper, we use an ensembled semi-supervised approach to determine the credibility of Reddit posts by estimating their reputation score to ensure the validity of information ingested by AI systems.  
We demonstrate our approach in the cybersecurity domain, where security analysts utilize these systems to determine possible threats by analyzing the data scattered on social media websites, forums, blogs, etc.
\end{abstract}

\section{INTRODUCTION}
Artificial Intelligence (AI) is widely utilized in diverse domains of industries like, finance, cars, cybersecurity, education, etc. AI systems are `trained' to learn complex problems and automate them for a larger scale. These systems need training data which is generally extracted and represented in a form that best suits the problem. One such source of data is {\em overt} or in a traditional cybersecurity sense, a part of the `Open-source Intelligence' (OSINT) \cite{doi:10.1080/08850609508435298}. OSINT includes data from sources such as newspapers, blogs, discussion groups, radio, social media websites, press conferences, journals, technical reports, etc. Online Social Media (OSM) is an OSINT source providing data that is ingested by AI tools working in various fields like finance \cite{jin2017tracking} and cybersecurity \cite{mittal2016cybertwitter}. Some of the most commonly used OSM are Twitter, Reddit\footnote{\url{https://www.twitter.com}, \url{https://www.reddit.com}}, etc. 

In cybersecurity, threat intelligence can be mined using {\em traditional} sources like NIST's National Vulnerability Database (NVD)\footnote{\url{https://nvd.nist.gov/}}, United States Computer Emergency Readiness Team (US-CERT)\footnote{\url{https://www.us-cert.gov/}}, etc. Other sources which are more {\em non-traditional} are, Twitter, Reddit, blogs, and news. Non-traditional sources are faster than the traditional ones. There is a significant gap between initial vulnerability announcement and NVD release \cite{attack2017register}. Vulnerability threat intelligence appears first on non-traditional sources \cite{vul2017register}. Mining non-traditional sources is becoming really important. In our previous work, we have developed \emph{CyberTwitter} \cite{mittal2016cybertwitter} and \emph{Cyber-All-Intel} \cite{mittal2017thinking} systems that mines threat intelligence from various OSINT sources. The systems then represent cybersecurity intelligence in knowledge graphs and vector spaces so it can be used by artificial intelligence based cyber-defense systems.

A new class of `Analyst Augmentation Systems' are being developed. More security analysts use these Artificial Intelligence based organizational cyber-defense systems to listen for threat intelligence mined from traditional and non-traditional sources, identify new vulnerabilities, analyze network and endpoint activity, find evidence of preplanned attacks and hints of data breaches.

The very `open' nature of these OSINT sources is its boon and its bane. These open channels are susceptible to `poisoning attacks' by a malicious entity. In a recent poisoning attack on Twitter, billions were wiped of the US Stock market when an Associated Press tweeted that then President, Barak Obama, had been injured in bomb blasts at the White House. This hack into Associated Press's Twitter account sent Dow Jones plunging 145 points in two minutes and S\&P 500 by nearly 1\% thereby incurring a loss of \${136.5 billion} \cite{6}. Data from OSM is vulnerable to misinformation in the form of hoaxes, fake images, videos, and rumors. 
This traditionally constitutes as fake news \cite{lazer2018science}. Several of these fake news incidents have caused a loss of money, reputation, infrastructure and in certain cases, threat to human lives. 
\begin{figure}[ht]
    \centering
        \includegraphics[scale=0.25]{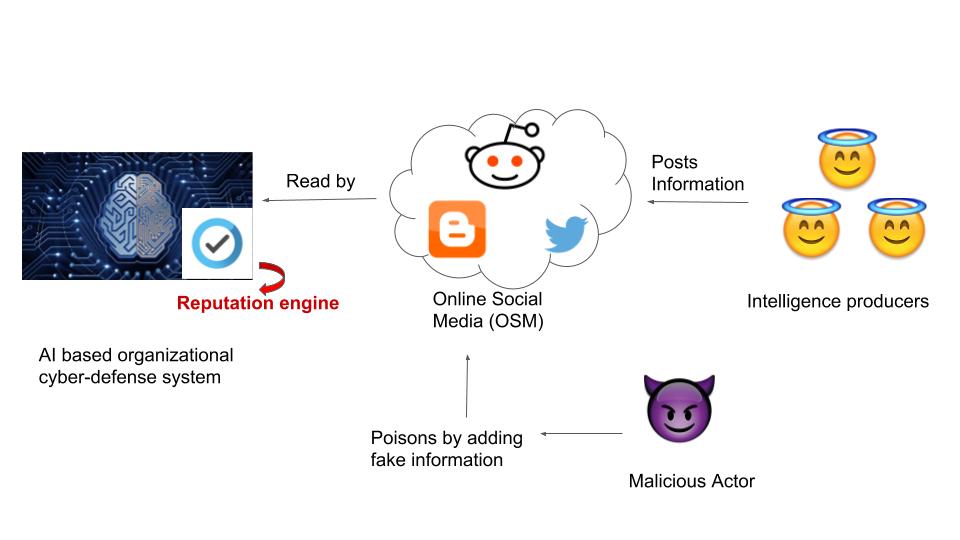}
        \caption{Attack scenario and proposed defense. Fake or contradictory information added by Attacker is verified using a reputation engine.}
        \label{fig:propsys}
\end{figure}

Increasing adoption of these non-traditional sources in AI cyber defense systems have created a potential attack surface. In an ideal environment, everything available on non-traditional intelligence sources will be credible and security analysts will mine information from these sources, to identify new vulnerabilities and then train their systems \cite{vul2017register}. However, in a realistic world, attackers, want to get past these AI cyber defenses by spreading misinformation. They can `poison' the data by adding incorrect information, for example, an attacker might spread the information that there exists a buffer overflow in Mozilla Firefox, this might trigger a policy change directive by a defensive AI. An attacker might use this as a diversionary tactic against the AI. Figure \ref{fig:propsys} explains this attack scenario.

They can also put in contradicting information about a valid threat intelligence. for example, an attacker might publish the information that a buffer overflow vulnerability exists in software MySQL, wherein MySQL has a SQL injection vulnerability. In this case the contradicting information will make the AI system more susceptible to an attack. The AI system will devote organizational resources like, analyst time, policy updates, network and endpoint defensive rule updates, etc. to protect against contradictory intelligence. A special case of contradictory information is `negative' intelligence. For example, in such a scenario the attacker publishes information that there is no SQL injection vulnerability or a valid vulnerability intelligence is false. This will reduce the confidence, that an AI system will place on a valid intelligence. 

OSINT as such, if consumed by the AI cyber defense system, can help the attacker evade various security measures thereby putting the organization at risk. Figure \ref{fig:propsys} explains this attack scenario and proposed defense. In this paper, we ensemble a SVM and an embedding model to build a reputation engine that checks the credibility of gathered intelligence information before it is consumed by the defensive AI. The reputation engine calculates a reputation score for each post and based on the generated score, recommends it for consumption. We use vector embeddings generated using the broad cybersecurity corpus created for the Cyber-All-Intel system \cite{mittal2017thinking} (See Section \ref{cluster}). The reputation score can be used by the AI system and the security analyst to threshold and control the level of trust in the incoming intelligence. The SVM classifier is used to classify posts as `credible' and `non-credible' using Reddit posts and `Redditor' features. More details about our proposed engine are described in Section \ref{method}.

The remaining paper is organized as follows: Section \ref{background} describes the background and the related work. Section \ref{method} discusses our methodology including data collection, vector generation, SVM classification, ensemble generation and reputation score calculation. Section \ref{results} summarizes our results. We conclude in Section \ref{conclusion}.

\section{RELATED WORK \& BACKGROUND}

In this section we discuss the background and the related work in the field of cybersecurity, artificial intelligence, credibility, and provenance. 

\subsection{AI for Cybersecurity}

Various representation techniques like knowledge graphs and vector space embeddings have been used to provide AI systems with knowledge about cybersecurity. 

Knowledge graphs have been used in cybersecurity to combine data and information from multiple sources which then aids a security analyst in her day to day operations. Various ontology based intrusion detection systems \cite{Undercoffer2003b,kandefer2007symbolic,takahashi2010ontological,takahashi2010building} have been put forth by researchers. These systems depend on a data repository of system vulnerabilities and threats 
\cite{Joshi-ICSC-2013,mittal2016cybertwitter}. These repositories are stored as RDF\footnote{\url{https://www.w3.org/RDF/}} linked data created from vulnerability descriptions collected from the National Vulnerability Database, Twitter, etc. Joshi et al. \cite{Joshi-ICSC-2013} extract information on cybersecurity-related entities, concepts and relations which is then represented using custom ontologies for the cybersecurity domain and mapped to objects in the DBpedia knowledge base \cite{auer2007dbpedia} using DBpedia Spotlight \cite{mendes2011dbpedia}. CyberTwitter \cite{mittal2016cybertwitter}, a framework to automatically issue cybersecurity vulnerability alerts to users. CyberTwitter converts vulnerability intelligence from tweets to RDF and uses the UCO ontology (Unified Cybersecurity Ontology) \cite{syed2015uco} to provide their system with cybersecurity domain information. Mittal et al. have also created \emph{Cyber-All-Intel} where they have used multiple knowledge representations to store threat intelligence \cite{mittal2017thinking}. 

Systems like the one proposed in \cite{mittal2016cybertwitter,mittal2017thinking} that extracts information from OSINT are susceptible to various attacks. For example, a possible attack on our proposed system is that the attacker can `poison' data sourced through multiple sources like Blogs, Social media, Dark Web, etc. i.e. an attacker can spread the information that there is a vulnerability in Microsoft Windows, even when such a vulnerability does not exist. In such a scenario we need to ensure that the credibility of the information being added to our cybersecurity corpus is checked by a reputation engine as discussed in Section \ref{method}. 

\subsection{Attacks on AI}
AI systems are susceptible to threats posed by malicious inputs \cite{vul2017register}, \cite{attack2017register}. Stevens et al. \cite{Stevens2016SummoningDT} describes how malicious inputs exploiting implementation bugs in ML algorithms pose a threat to organizations. They have defined the term `poisoning attacks' and `evasion attacks' as an exploit targeting the training and testing phase respectively. They used a semi-automated technique, called steered fuzzing to explore the attack surface and calculate the magnitude of the threat. 

\subsection{Credibility of Intelligence}

Several models or tools have been developed over the past to identify `poisoning' of data in a generic sense. Our work aims at creating a credibility system for Threat Intelligence.

One such system is `TweetCred' \cite{gupta2014tweetcred}, that assigns a `credibility score' to every tweet to identify fake tweets and thereby providing valuable information during crisis to emergency responders and the public. It was devised to identify the credibility of tweets motivated by false tweets published during `high impact events'. 
Rakib et al. used word embeddings on Reddit database based on word2vec skip-gram model to train a random forest classifier to identify cyberbully comments \cite{10.1007/978-3-319-75417-8_17}.
We build upon these systems to assign a ’reputation score’ for threat intelligence mined from Reddit. 
On Reddit, each account is associated with some meta-data which is the user profile information, the posts written using that account and the network information which comprises of its connections with other user accounts. We use these features and other latent semantic models to compute the reputation score (See Section \ref{method}).  

\section{METHODOLOGY}
\label{method}

\label{background}
\begin{figure}[ht!]
    \centering
        \includegraphics[scale=0.26]{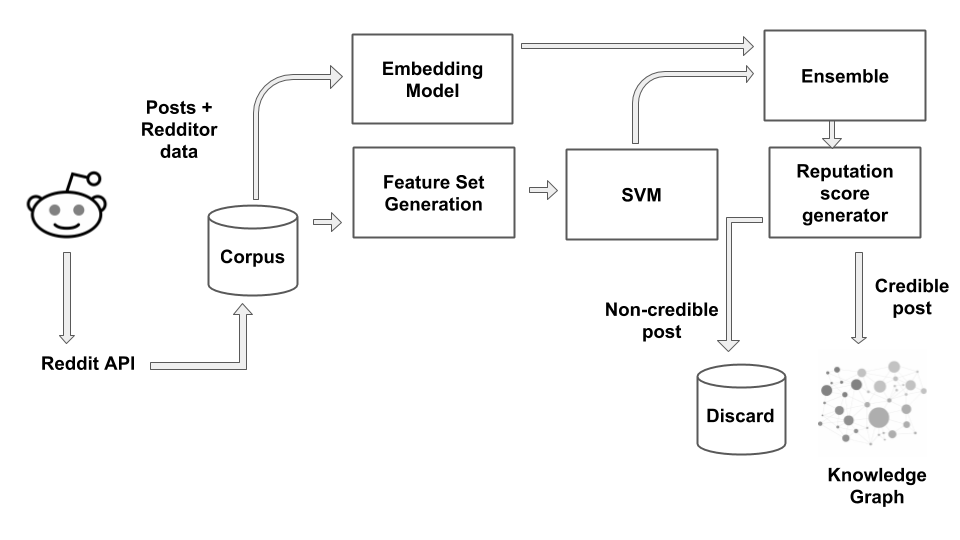}
        \caption{Architecture of our methodology and analysis.}
        \label{fig:arch}
\end{figure}

In this section, we describe the overall architecture (See Figure \ref{fig:arch}) of our proposed system that includes a reputation engine to calculate the reputation score for each post. The system was created by generating vector embeddings for Reddit posts. We use a semi-supervised learning algorithm, where we have a lot of unlabeled data with a small quantity of labeled data. Our approach leverages the cluster and continuity assumptions which are a universally accepted parts of various semi-supervised learning algorithms \cite{chapellesemi}. The reputation score is generated using the distance of a post's embeddings from `credible' and `non-credible' clusters. The SVM classifier is used to classify posts as `credible' and `non-credible' using Reddit posts and `Redditor' features.

\subsection{Data Collection}
Reddit is a social news aggregation, web content rating, and discussion website with over 230 million users \cite{weninger2013exploration}. The data is segregated into different tabs within each subreddit. 
We collected data from Reddit using the 
Reddit API \footnote{\url{https://www.reddit.com/dev/api/}} 
The API gives an instance of Reddit that can be used to obtain all the `hot', `new', `controversial', `gilded' or `top submission' instances. It also provides the data on submitter of the post (also termed as a `Redditor') and various comments. We collected 14,500 posts corresponding to several cybersecurity subreddits like: cybersecurity, malware, cryptography, cyberlaw and cybersecurityfans, etc.

\subsection{Labeled dataset generation}\label{feature}
Human annotators were used to obtain the ground truth for our experiments. 
From the 14,500 posts, we randomly picked a sample of 2000 posts for annotation. We provided the annotators the definition of credibility and asked them to classify the posts into two classes: `credible' or `non-credible'. Annotators were given added information like referred Common Vulnerabilities and Exposures (CVE) database entries and links to verified news websites like, The Washington Post \cite{washingtonpost}, BBC \cite{bbc}, The Guardian \cite{guardian}, CNN \cite{cnn}, Reuters \cite{reuters}, etc. or cybersecurity sources like HackerNews \cite{hackernews}, Krebs on Security \cite{krebs}, Microsoft \cite{microsoft}, etc. 
We annotated cyber Reddit posts with the help of 5 graduate students with specialization in cybersecurity to obtain the ground truth regarding the credibility of posts. Each post was annotated by 3 annotators. We calculated the Cohen's Kappa score to check the reliability of the results obtained by annotation. Each post was annotated by at-least 3 annotators to get a good inter annotator agreement. The inter-annotator agreement for all posts was calculated and posts with score $>$ 0.66 were kept. We obtained 1206 posts that served as ground truth with 953 posts entitled as `credible' and 253 as `non-credible'. 


\subsection{Reddit Post Vectors}\label{cluster}

In our supervised model, we also incorporated vector projections of the post to help classify them as credible or non-credible. We create embeddings for the posts in which each post is modeled as a `bag of words' and represented as a sum of it's word embeddings. All the word vectors are summed up to get the total vector value of the post. The word embeddings were taken from the model created by Mittal et al. for their \emph{Cyber-All-Intel} system \cite{mittal2017thinking}. To create the Reddit post embeddings we took the following steps:
\begin{enumerate}
\item Generate individual cybersecurity word embeddings: We used a cybersecurity corpus collected from multiple OSINT sources like National Vulnerability Datasets, security bulletins, security blogs, Twitter, Reddit, etc. The text corpus and word embeddings were taken from the Cyber-All-Intel system \cite{mittal2017thinking}. Taking a corpus collected using different OSINT sources provides the system a more global view of the cybersecurity landscape.

\item Extract cybersecurity concepts and vulnerabilities present in Reddit posts: We use a Security and Vulnerability Concept Extractor (SVCE) to extract terms related to cybersecurity \cite{Lal-MS-2013,mittal2016cybertwitter}. The SVCE is able to tags every sentence with the following concepts: Means of an attack, Consequence of an attack, affected software, hardware and operating system, version numbers, network related terms, file names and other technical terms.

\item Creating Reddit post vectors: Once we get the output of the SVCE, we fetch the corresponding word embeddings from the word embedding model mentioned in Step 1. Each post is then represented as the sum of cybersecurity term vectors present in that post (This is a slight modification to Doc2Vec\footnote{\url{https://radimrehurek.com/gensim/models/doc2vec.html}}). In our implementation we do not include non-cybersecurity terms in the post representation as we empirically found that it adds noise to the system.

\end{enumerate}

Using the ground truth post's vectors we create 2 clusters: `credible' and `non-credible'. We use these to compute the reputation score. A visual representation has been shown in Figure \ref{fig:cluster}. We evaluate the quality of vectors generated in Section \ref{results}.



\subsection{SVM Classification}
\label{svm}

We trained a SVM classifier that we ensemble with the embedding model mentioned in Section \ref{cluster}. We begin by defining the feature set and then train a classifier using Support Vector Machine (SVM). The following are the features that we include in our model, these features have been collected using the Reddit API:
\begin{itemize}
\item \textit{Post features}: Length of a post, seconds passed since it was posted, downvotes, upvotes, score, number of comments, number of crossposts and Web of Trust (WOT)\footnote{\url{https://www.mywot.com}} values of URLs.
\item \textit{Redditor features}: Redditor\'s screen name length, seconds since the user registered, link\_karma, comment\_karma, verified user email, verified user, user is a moderator or not.
\end{itemize}

After training Linear SVC on the annotated 1206 posts, we obtained a learned model that classifies posts for credibility. We classified the posts into two classes `credible' and `non-credible'.
Next we discuss the ensemble model and the reputation score generation. We evaluate all three models in Section \ref{results}.

\subsection{Ensemble \& Reputation score generation}
\label{ensemble}

\begin{figure}[ht!]
    \centering
        \includegraphics[scale=0.42]{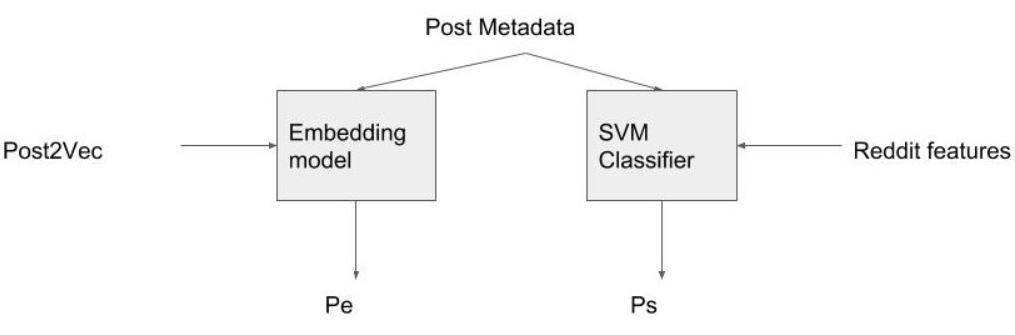}
        \caption{Ensembling: Stacking Embedding model with SVM Classification.}
        \label{fig:ensemble}
\end{figure}

In our system, we ensemble the two models: SVM classifier and vector embeddings to identify the validity of our posts. We use stacking method to improve the prediction of our system. Stacking models in parallel combines all the classifiers and creates a meta-classifier. In the first step, we utilize the vector embedding model as a base model that is trained on the complete training set of 1206 posts and the post vectors thus obtained as output are collectively used with other identified features for our meta-model, SVM classifier. Figure \ref{fig:ensemble}, gives details on how the two classifiers are combined. The embedding model produces a predictive measure $Pe$ and the SVM classifier produces $Ps$. The weighted sum of these two models gives the final prediction ($Pf$) for the credibility of a post. The weights $We$ and $Ws$ are learned experimentally. A system analyst can set a threshold for $Pf$. For our experiments we use $Pf > 0.6$ as credible.  

\[Pf = \frac{(We*Pe) + (Ws*Ps)}{(We + Ws)}\]

\[Pf, Ps, Pe \in [0, 1]\]

Next, we wish to create a quantifiable score which can be understood by both the AI system and the security analyst. We calculate the reputation score of a post by determining the distance of the post vector from the cluster centroids created in Section \ref{cluster} and the output of the Ensemble unit. The score $s_{c}$ is calculated with respect to the distance from `credible' cluster ($d_{c}$) and the distance from the `non-credible' cluster ($d_{i}$) as:

\[s_{c} = 1 - \frac{d_{c}}{d_{c}+d_{i}}\]

We use both the ensembled SVM classifier along with the vector embeddings to predict if a post is `credible' or `non-credible' and it's reputation score. We also identify the features that serve as strong indicators of credibility for classification by determining the weighted classifier coefficients. We discuss the same in Section \ref{results}.



\section{RESULTS \& EVALUATION}\label{results}

This section describes the results obtained on classifying posts using the ensembled model. We first evaluate the vector embedding model followed by the SVM model. We also discuss the features that turned out to be strong indicators of credibility.

\subsection{Quality of Vector Embeddings}

In Section \ref{cluster} we discuss our Reddit post vector generation. 
The post vectors were evaluated manually, by randomly taking 50 posts and then analyzing 5 similar posts retrieved, for each of the 50 using the embedding model. The annotations were done by 2 annotators who evaluated if the retrieved posts were similar to the input post. The mean average precision recorded for the same was 0.59. We would like to point out the fact that this evaluation scheme is really expensive. The results of the credible and non-credible classification using just the embedding model has been shown in Table \ref{5050derive}.

\subsection{SVM Classification Analysis}
We used the Support Vector Machine (SVM) over the selected features described in Section \ref{svm} to estimate the credibility of the posts. After training SVM on the annotated 1206 posts, we obtained a learned model that classifies posts for credibility. 
The results of the credible and non-credible classification using just the SVM model has been shown in Table \ref{5050derive}. 

As a result of our analysis, we identified the following features as strong indicators of credibility: the time at which the post was submitted, the Web Of Trust (WOT) score of the URL in the post, post's length and `Redditor' features such as link and comment karma. 
High value of the WOT score of the post URL indicates high credibility of the URL from which the data is extracted. High WOT score websites are observed to be the verified news websites like The Washington Post \cite{washingtonpost}, BBC \cite{bbc}, The Guardian \cite{guardian}, CNN \cite{cnn}, Reuters \cite{reuters}, etc. or cybersecurity sources like HackerNews \cite{hackernews}, Krebs on Security \cite{krebs}, Microsoft \cite{microsoft}, etc. Thus, presence of a URL in a post showed a strong positive correlation with credibility. The length and submission time of the post and also suggested high credibility of the post; informative and older posts seem to be credible. Some other important indicators were Redditor's link and comment karma. A link karma shows the number of links posted by a `Redditor' and comment karma exhibits the number of posted comments and upvoted by other `Redditors'. `Redditors' who have been active and posted more comments and links are trusted and usually post credible posts.
Hence, the post attributes and `Redditor' features played an important role in determining credibility.

\subsection{Ensembled Model}

For the ensembled model, we take the output of the embedding model and the SVM to create a stacked meta-classifier (see Section \ref{ensemble}). On evaluating the ensembled model, we get a ten-fold cross validated accuracy of  71.54\%.

\begin{table}[!h]
\centering
\begin{tabular}{| c | c | c|}
\hline
& {\bf \small True Positive} & \bf\small{False Positive} \\ \hline
\bf\small{False Negative} & 188 & 67 \\ \hline
\bf\small{True Negative} & 77 & 174  \\ \hline
\end{tabular}
\caption{Confusion matrix for balanced set of `credible' and `non-credible' posts for the Ensembled model}
\label{5050}
\end{table}
\begin{table}[!h]
\centering
\begin{tabular}{| c | c | c | c |}
\hline
\bf\small{Metrics} & \bf\small{Ensembled} & \bf\small{Embedding} & \bf\small{SVM}\\\hline
Accuracy & 71.541\% & 66.919\% & 58.02\% \\ \hline
Precision & 0.72199 & 0.66900 & 0.57\\ \hline
Recall & 0.69323 & 0.62 & 0.554\\ \hline
True Negative Rate & 0.73725 & 0.6832 & 0.604\\ \hline
False Positive Rate & 0.26274 & 0.3111 & 0.395\\ \hline
F1 Score & 0.70732 & 0.6525 & 0.575\\ \hline
\end{tabular}
\caption{Confusion matrix and derived metrics for a balanced set. The weights for the ensembled model were $We$ = 0.58 and $Ws$ = 0.47. We take $Pf > 0.6$ as credible}
\label{5050derive}
\end{table}

Table \ref{5050} describes the confusion matrix obtained for the predicted posts. 
Thus, our analysis for credibility predicted results with an accuracy of 71.541\%. We also computed other derived metrics (Table \ref{5050derive}) for a balanced set of `credible' and `non-credible' posts.

\begin{table}[!h]
\centering
\begin{tabular}{| p{4.2cm} | p{1.1cm} | p{1.1cm} | p{0.7cm} |}
\hline
\bf\small{Post} & \bf\small{Distance from Credible cluster} & \bf\small{Distance from Non-credible cluster} & \bf\small{Rep. score}\\ \hline
I have just tried it and this exploit just works !!! Joomla powered websites that have ``Joomanager 2.0.0'' & 0.00697 & 0.02646 & 0.791\\ \hline
Turns out the Verge fiasco is worse than thought. Devs now having to issue new wallets having accidentally hardforked their own currency trying to fix the attack. Popcorn, salt and GODL overflowing &  0.02986 & 0.00343 & 0.103\\ \hline
\end{tabular}
\caption{Distance of post's vector from centroid of two clusters.}
\label{table4}
\end{table}
\begin{figure}[ht!]
    \centering
        \includegraphics[scale=0.7]{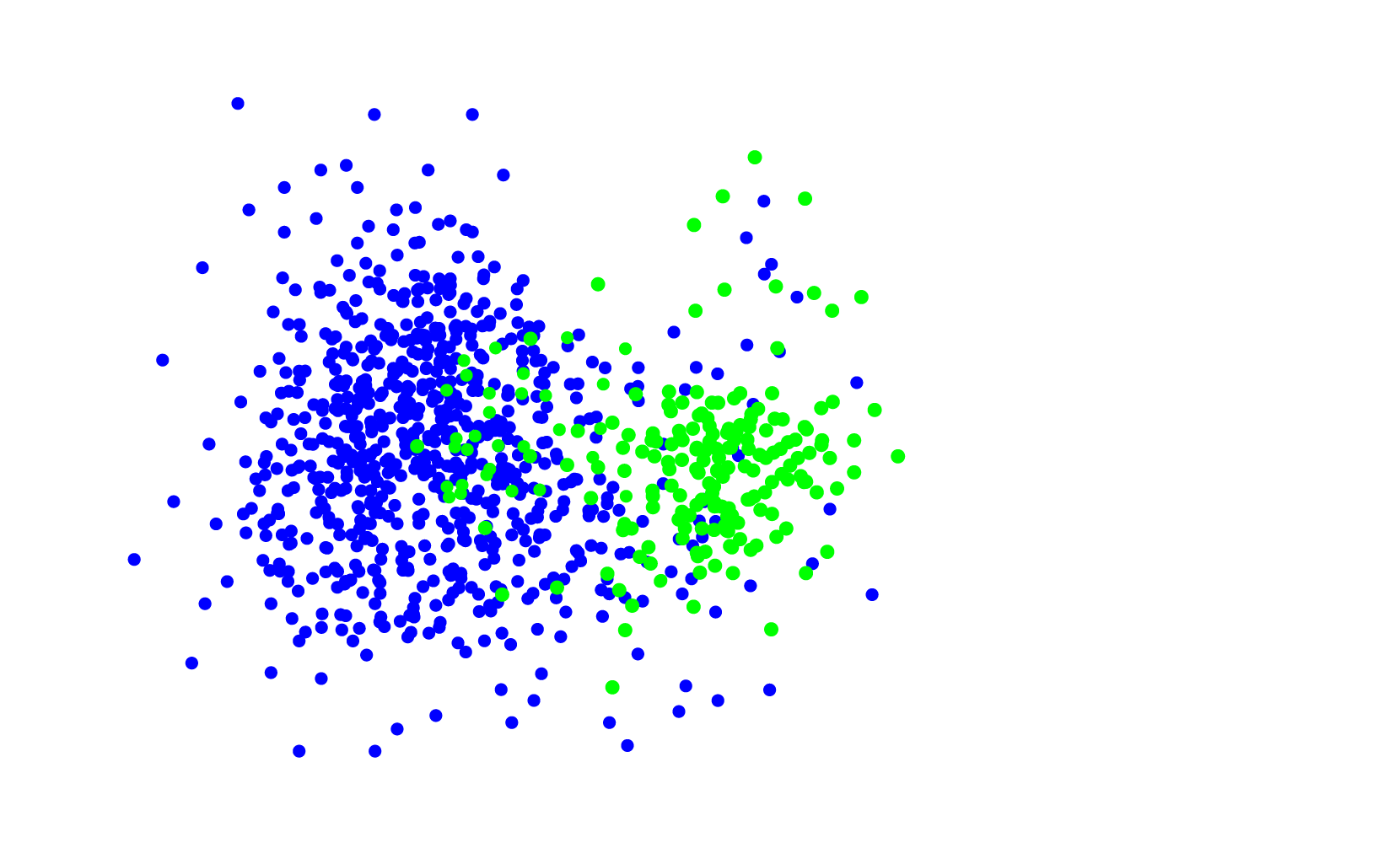}
        \caption{Visualization of post clusters using t-SNE. Blue cluster represents `credible' posts and green represents `non-credible' annotated posts.}
        \label{fig:cluster}
\end{figure}
We also calculated the reputation score of the posts using their relative distances from the credible and non-credible clusters obtained from ground truth post vectors. Figure \ref{fig:cluster}, shows that posts identified as `credible' by classification tend to lie in close proximity of credible cluster and `non-credible' posts lie close to non-credible cluster. The distance from the centroids of the two clusters for two sample posts is listed in Table \ref{table4}. 
The first post was identified as `credible' by our analysis and was closer to the credible cluster and the second post was closer to the incredible cluster and identified as `non-credible'. The minimum of the distances of the post's vector from the centroids of the two clusters gave its reputation score. Hence, post 1 had a reputation score of 0.791 and post 2 received a score of 0.103. 



\section{CONCLUSION AND FUTURE WORK}
\label{conclusion}
With the rise in use of online social media (OSM) and data analysis, AI systems have been widely used for predictive analysis. The information extracted from these sources is prone to poisoning. 

In the domain of cybersecurity, OSMs have become a source of threat intelligence gathering. This threat intelligence is usually ingested by various cyber-defense systems. 
The AI systems are exposed to poisoning attacks if we do not perform a credibility check before an intelligence is ingested by a cyber-defense AI. In this paper, we create a reputation engine to calculate the credibility of the threat intelligence. We have evaluated the credibility of Reddit posts that belong to cybersecurity, cyber, malware, cryptocurrency, cryptomarkets, cyberlaw, etc. subreddits. We created a ensembled semi-supervised model to calculate the reputation of Reddit posts, related to cybersecurity.   We ensembled an embedding model and a SVM model. Ground truth was established using manual annotation of posts that were used to train our model and predict the credibility of posts. We classified the posts as `credible' or `non-credible' with an accuracy of 71.73\%. 
The reputation score of the posts was evaluated based on the distance of the post vector from the centroids of the clusters plotted for posts in a vector space. We established that both content and `Redditor' features play a vital role in determining the credibility of a Reddit post. 


In the future, we would establish more ground truth data for our analysis to further improve the accuracy of our system. We have used an ensembled semi-supervised approach; such an approach usually yields better results with more annotated data. Getting more annotated data is expensive, access to more ground truth will help in better evaluation and training. Also, we would like to incorporate other online social networks like Quora \cite{quora}, Twitter, dark web, etc. as they are widely used for discussions about cybersecurity threats and vulnerabilities. We would also like to include a validation scheme where vendors can put their threat intelligence as verified. Vendors can tag their intelligence as verified in the form of a tag or an attribute. We would also like to develop a User Interface or a tag with each post displaying its reputation score or ask for a feedback if the user does not agree with the calculated score.
\section*{ACKNOWLEDGEMENT}
\label{ack}

The work was partially supported by a gift from IBM Research, USA.


\bibliographystyle{plain}
\bibliography{Bibliography,phd}
\end{document}